\RequirePackage{ifpdf}
\ifpdf
	\documentclass[twocolumn, showkeys, letterpaper, pdftex]{revtex4}
\else
	\documentclass[twocolumn, showkeys, letterpaper, dvips]{revtex4}
\fi

\usepackage{amsmath}

\ifpdf
  \usepackage[pdftex]{graphicx}
  \usepackage{subfigure}
  \graphicspath{{pdf-figures/}}
  \usepackage[%
    pdftex,%
    pdftitle = {{Strained substrates, composition inhomogeneities, grain shape and the agglomeration of germanosilicide thin films}},%
	pdfsubject = {{The grains of metal germanosilicide films formed on silicon-germanium(001) are polygonal, Ge composition is not homogeneous, and substrates are typically strained. These effects can explain why these films agglomerate at lower temperature than silicides.}},%
	pdfauthor = {{Mathieu Bouville}},%
	pdfkeywords = {{silicide, germanide, agglomeration, segregation, dealloying, grain-boundary grooving, faceting, thermodynamics, theory}},%
  ]{hyperref}
\else
  \usepackage[dvips]{graphicx}
  \usepackage{subfigure}
  \usepackage[
	dvips,
    pdftitle = {{Strained substrates, composition inhomogeneities, grain shape and the agglomeration of germanosilicide thin films}},%
	pdfsubject = {{The grains of metal germanosilicide films formed on silicon-germanium(001) are polygonal, Ge composition is not homogeneous, and substrates are typically strained. These effects can explain why these films agglomerate at lower temperature than silicides.}},%
	pdfauthor = {{Mathieu Bouville}},%
	pdfkeywords = {{silicide, germanide, agglomeration, segregation, dealloying, grain-boundary grooving, faceting, thermodynamics, theory}},%
	letterpaper,%
]{hyperref}
\fi

\newcommand{\GB}{\ensuremath{_\mathrm{gb}}}
\newcommand{\I}{\ensuremath{_\mathrm{i}}}
\newcommand{\SiGe}{\ensuremath{_\text{xz}}}
\newcommand{\NiSi}{\ensuremath{_\mathrm{MSi}}}
\newcommand{\NiGe}{\ensuremath{_\mathrm{MGe}}}
\newcommand{\Hform}{\ensuremath{H}}
\newcommand{\Si}{\ensuremath{_\mathrm{Si}}}
\newcommand{\Ge}{\ensuremath{_\mathrm{Ge}}}
\newcommand{\IV}{\ensuremath{_\mathrm{IV}}}
\newcommand{\MIV}{\ensuremath{_\text{M--IV}}}
\newcommand{\x}{\ensuremath{_\mathrm{x}}}
\newcommand{\z}{\ensuremath{_\mathrm{z}}}
\renewcommand{\u}{\ensuremath{_\mathrm{u}}}
\newcommand{\el}{\ensuremath{\mathcal{E}}}
\newcommand{\A}{\ensuremath{\mathcal{A}}}
\newcommand{\eps}{\varepsilon}
\newcommand{\f}{\ensuremath{_\mathrm{f}}}
\newcommand{\s}{\ensuremath{_\mathrm{s}}}
\newcommand{\facet}{\ensuremath{_\mathrm{poly}}}
\newcommand{\iso}{\ensuremath{_\mathrm{circ}}}
\newcommand{\kT}{\ensuremath{R\,T}}
\newcommand{\dz}{\ensuremath{\delta \alpha}}

\begin{document}

\title{Strained substrates, composition inhomogeneities, grain shape\\and the agglomeration of germano\-silicide thin films}

\author{Mathieu Bouville}
\email{m-bouville@imre.a-star.edu.sg}
\affiliation{Institute of Materials Research and Engineering, Singapore 117602}
\date{\today}

\begin{abstract}
Germano\-silicide thin films are quite different from silicides and germanides. The germanium composition is not homogeneous, grains have a different shape, and the substrate is generally strained. This affects grain boundary grooving and favors agglomeration of germano\-silicide films.
Our thermodynamics model shows that the equilibrium Ge composition of germano\-silicide films formed on strained substrates is inhomogeneous and these films are more likely to agglomerate than those formed on relaxed substrates, in agreement with experiments.
Grain shape too can affect agglomeration: poly\-gonal films, such as germanosilicides, are more likely to agglomerate than films with rounded grains (silicides, germanides).
\end{abstract}


\maketitle

Metal germano\-silicide thin films formed on Si$_{1-x}$Ge$_x$ are attractive to the semiconductor industry for use in advanced metal--oxide--semiconductor field-effect transistors (MOSFETs) because of the higher carrier mobility of germanium compared to silicon. Also, in the case of nickel, the difficulty to form NiGe$_2$ pushes the formation of the high-resistivity disilicide phase to higher temperature~\cite{Seger-JAP-04}.
On the downside germano\-silicides tend to agglomerate at lower temperature than silicides \cite{Seger-JAP-04, Ok-SST-04, Zhao-microelec-04}. MSi$_{1-u}$Ge$_u$ materials systems (where M is a metal) are very different from MSi and MGe: (i) the germanium content in the film is not the same as that in the substrate and the composition of the substrate is not homo\-geneous~\cite{Seger-JAP-04, Yao-05, zhang-microelec-03, Ok-SST-04, Yao-essl-04}, (ii) the interface between a MSi$_{1-x}$Ge$_x$ grain and the substrate may not be an arc of a circle as in MSi or MGe~\cite{Yao-05, Seger-JAP-04, Chen-micron-02, Zhao-microelec-04, Yao-essl-04}, and (iii) the substrate is typically under compression whereas silicides are generally formed on relaxed Si substrates.

\begin{figure}
\centering
    \includegraphics[width=8.5cm]{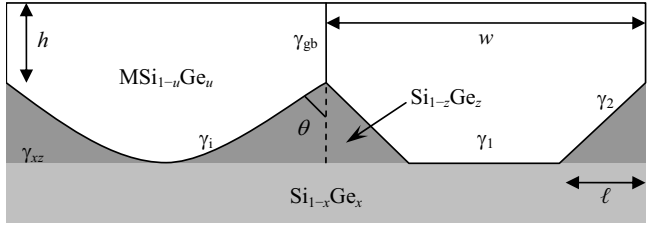}
\caption{\label{fig-schematic}Rounded grain (left) and polygonal grain (right) of MSi$_{1-u}$Ge$_u$ on a Si$_{1-x}$Ge$_x$(001) substrate in which there exists a region with a different composition, Si$_{1-z}$Ge$_z$
. $\gamma\GB$ is the grain boundary energy. $\gamma\I$, $\gamma_1$, and $\gamma_2$ are the film--substrate interface energies and $\gamma\SiGe$ is the Si$_{1-z}$Ge$_z$--Si$_{1-x}$Ge$_x$ interface energy.}
\end{figure}

During annealing, the interface morphology evolves due to thermal grooving and films agglomerate when the interface reaches the surface.
This corresponds to the height of the grain boundary, $h$ in Fig.~\ref{fig-schematic}, reaching zero. $h$ thus determines the stability of the film: if $h$ is always positive the system cannot agglomerate whereas it can agglomerate if $h$ can be equal to zero.
In this letter, we study the effects of substrate strain and of grain geometry on film stability. We first study as a function of substrate strain a system in which the film--substrate interface is an arc of a circle. We then compare films with rounded and faceted interfaces formed on relaxed substrates.

The system is two-dimensional and, following experimental observations in NiSi$_{1-u}$Ge$_u$/\linebreak[3]Si$_{1-x}$Ge$_x$(001)~\cite{Seger-JAP-04, Yao-05, zhang-microelec-03, Ok-SST-04, Yao-essl-04}, we account for three different germanium compositions: $x$ in the bulk of the substrate, $z$ at the groove, and $u$ in the film (Fig.~\ref{fig-schematic}).
We determine the equilibrium values of these compositions and of $h$ and $\theta$ by minimizing the system free energy,
\begin{align}
G=	&\left[ \Hform\NiSi + u \Delta H\MIV +\textstyle\frac{1}{2}\kT S(u) +\el\u\right]w \,h\f\nonumber\\
	&+[\Hform\Si +x \Delta H\IV +\kT \,S(x) +\el\x] \left( w \,h\s - \A\z \right)\nonumber\\
	&+[\Hform\Si +z \Delta H\IV +\kT \,S(z) +\el\z] \,\A\z \nonumber\\
	&+\gamma\GB\,w + \gamma\I \, w(\pi/2-\theta)/\!\cos\theta +\gamma\SiGe \, w.
\end{align}
\noindent $h\f$ and $h\s$ are the initial film and substrate thicknesses respectively, $R$ is the gas constant, $T$ is the temperature, and $S(a) \equiv a \ln a +(1-a) \ln(1-a)$. $\Hform\NiSi$, $\Hform\NiGe$, $\Hform\Si$, and $\Hform\Ge$ are formation enthalpies; $\Delta H\IV \equiv \Hform\Ge-\Hform\Si$ and $\Delta H\MIV \equiv \Hform\NiGe-\Hform\NiSi$. $\A\z$ is the surface area of the Si$_{1-z}$Ge$_z$ region.
\noindent The first three terms of $G$ are the free energies of MSi$_{1-u}$Ge$_u$, Si$_{1-x}$Ge$_x$, and Si$_{1-z}$Ge$_z$ respectively. The other terms are grain boundary and interface energies. 

$\el\x = \frac{1}{2} \alpha\x C \eps^2 (x-y)^2$, $\el\z = \frac{1}{2} \alpha\z C \eps^2 (z-y)^2$, and $\el\u$ are the elastic energies of Si$_{1-x}$Ge$_x$, Si$_{1-z}$Ge$_z$, and MSi$_{1-u}$Ge$_u$. $C$ is an elastic constant (independent of composition) and $\eps$ is the Ge--Si lattice mismatch. $\alpha\x$ and $\alpha\z$ are dimensionless geometric factors; 
as the Si$_{1-z}$Ge$_z$ region is geometrically less constrained than the Si$_{1-x}$Ge$_x$ region relaxation is possible and $\alpha\z<\alpha\x$. $y$ is the composition of a {\itshape relaxed} Si$_{1-y}$Ge$_y$ substrate which would have the same lattice parameter as the actual substrate. If the substrate is unstrained then $y=x_0$, the nominal Ge composition of the substrate. If the substrate is a Si$_{1-x}$Ge$_x$ film grown epitaxially on a Si wafer, $y=0$ as the lattice parameter of the substrate is that of silicon.

There are five unknowns ---three compositions ($u$, $x$, and $z$) and two geometric parameters ($\theta$ and $h$)--- 
and two relations between them (mass conservation of metal and group-IV species).
The free energy can therefore be written as a function of $u$, $z$, and $\theta$ and the equilibrium is given by $\partial G / \partial u = \partial G / \partial z = \partial G / \partial \theta = 0$. Interface energies \emph{a priori} depend on $u$, $z$, and $\theta$. However the details of these dependences are not known. We will therefore focus on the elastic energy, neglecting the derivatives of the interface energies.

The equilibrium for $u$ is
\begin{align}
	&\frac{u}{1-u} =  \frac{x}{1-x} \times \nonumber\\
	&\exp \left(\!-\frac{2\Delta H\MIV - \Delta H\IV - 
{\partial \el\x}/{\partial x}
+2{\partial \el\u}/{\partial u} }{\kT} \right)\!\!.
	\label{eq-u_x}
\end{align}
\noindent Since formation enthalpies are typically larger than elastic energies the value of $u$ is controlled by $2\Delta H\MIV - \Delta H\IV$. As this term is positive~\cite{zhang-microelec-03}, $u<x$. This is consistent with experimental observations of germanium depletion in the film~\cite{Seger-JAP-04, Yao-05, zhang-microelec-03, Ok-SST-04, Yao-essl-04}. If the elastic energies are omitted, Eq.~(\ref{eq-u_x}) simplifies to Eq.~(4) in Ref.~\onlinecite{zhang-microelec-03}. 
On a compressive substrate ${\partial \el\x}/{\partial x}$ is positive and $u$ increases: because a higher germanium composition in the substrate would raise the strain and the elastic energy, compressive substrates  reduce Ge out-diffusion.

To second order in $x-y$, the derivative of the free energy with respect to $z$ gives
\begin{equation}
z \approx x +\dz\,(x-y) +
\frac{\frac{1}{2}-x}{x(1-x)} \frac{\dz^2}{1+\alpha\z/\psi} (x-y)^2\!,
	\label{eq-z}
\end{equation}
\noindent where 
\begin{equation*}
	\psi = \frac{\kT}{C\eps^2}\frac{1}{x(1-x)}
		\text{\quad and \quad}
	\dz = \frac{\alpha\x-\alpha\z}{\alpha\z +\psi}.
\end{equation*}
\noindent If $\psi$ is small elasticity dominates and if it is large entropy dominates. 
The compositions $u$, $x$, and $z$ are obtained from Eqs.~(\ref{eq-u_x}) and~(\ref{eq-z}) and mass conservation,
\begin{equation}
	\frac{1}{2}w \,h\f \,u + \left( w \,h\s - \A\z \right) x + \A\z \, z = w \left(\frac{1}{2}h\f+h\s\right) x_0.
	\label{mass-cons}
\end{equation}
\noindent Here $x_0$ is the initial Ge composition. 

If the substrate is relaxed then $x=y=x_0$ and there is no segregation, $z=x$. If the Si$_{1-x}$Ge$_x$ substrate was grown on Si and did not relax then $y=0$ and there is segregation, $z>x$. This difference of Ge composition within the substrate has been observed experimentally~\cite{Seger-JAP-04, Yao-05, zhang-microelec-03, Ok-SST-04, Yao-essl-04}. This is similar to the case of Si$_{1-x}$Ge$_x$/Si where Ge atoms segregate to the islands where the system is not constrained geometrically and relaxation is possible. As Eqs.~(\ref{eq-u_x})--(\ref{mass-cons}) are independent of grain geometry (only the value for $\A\z$ in Eq.~(\ref{mass-cons}) depends on grain shape), so are the conclusions regarding Ge segregation.

In the case of a film with an interface which is an arc of a circle, 
to second order in $x-y$ the derivative of the free energy with respect to $\theta$ leads to
\begin{align}
	\cos\theta \approx \cos\theta_0 &+ \frac{C\eps^2(x-y)^2}{2\gamma\GB/w}\dz  (\alpha\x +\psi) \times \nonumber\\
		& \sin\theta_0 \frac{\pi/2-\theta_0-\cos\theta_0}{1 - (\pi/2-\theta_0)\tan\theta_0},
	\label{def-theta}
\end{align}
\noindent where
\begin{equation}
	\cos\theta_0=\frac{\gamma\GB}{2\gamma\I}.
	\label{def-theta_0}
\end{equation}
\noindent $\partial \el\u / \partial \theta$ and $\partial \alpha\z / \partial \theta$ have been neglected in order to simplify the equation. Without strain, one recovers the usual $\cos\theta={\gamma\GB}/{(2\gamma\I)}$~\cite{Bailey-PPSB-50, Mullins-JAP-57, Nolan-JAP-92}.

Thinner films and larger grains are known to be more prone to agglomeration. Under given experimental conditions, there is a minimum thickness for the continuous film to be stable~\cite{Nolan-JAP-92}. Stability can thus be determined from the aspect ratio of the initial grain, $h\f/w$. We call $r\iso$ the minimum value of $h\f/w$ to avoid agglomeration,
\begin{equation}
	r\iso=\frac{\pi-2\theta-\sin2\theta}{8\cos^2\theta}.
	\label{def-r_iso}
\end{equation}
\noindent $r\iso$ is such that at equilibrium $h=0$. From Eqs.~(\ref{def-theta}) and~(\ref{def-r_iso}), we find that $r\iso$ increases with $[\eps(x-y)]^2$: on a strained substrate the film needs to be thicker to avoid agglomerating, i.e.\ it is less stable. This is consistent with experiments which show that, compared to films formed on relaxed substrates, germanosilicides formed on compressive substrates agglomerate at lower temperature~\cite{Seger-JAP-04, Ok-SST-04, Zhao-microelec-04}. 

Equations~(\ref{eq-u_x}), (\ref{eq-z}) and~(\ref{def-theta}) were derived assuming equilibrium. However annealing ends or agglomeration occurs before equilibrium is reached. As the Si$_{1-z}$Ge$_z$ region is rather small and close to the grain boundary (fast-diffusion path) $z$ converges relatively fast to the value of Eq.~(\ref{eq-z}). Likewise the angle $\theta$ is established very quickly. The main difference for films out of equilibrium is that it takes time for the whole interface to become an arc of a circle (an arc of a circle gives a constant chemical potential) and $h$ is larger than its equilibrium value. Equation~(\ref{def-r_iso}) is thus an upper bound: films with a larger aspect ratio are stable but some films with a smaller value of $h\f/w$ may not agglomerate at short annealing times.

So far we assumed that the interface 
is an arc of a circle. However in germano\-silicide materials systems ---unlike in MSi/Si and MGe/Ge--- the grains may not be rounded~\cite{Yao-05, Seger-JAP-04, Chen-micron-02, Zhao-microelec-04, Yao-essl-04}. At the groove root, the interface tends to be flat which gives the grains a polygonal shape as illustrated in Fig.~\ref{fig-schematic}. 
This different shape of the grains may affect the agglomeration of thin films. 
The minimum value of the aspect ratio to avoid agglomeration of polygonal grains is
\begin{equation}
	r\facet=\frac{\ell}{w\,\tan\theta}\left( 1-\frac{\ell}{w} \right)\!\!.
	\label{def-r_facet}
\end{equation}
\noindent $\ell$ is obtained by minimizing $G$ as was done for a rounded film (we neglect strain to focus on geometry).

If the interface energy is isotropic, the angles of the triangular Si$_{1-z}$Ge$_z$ region are set by the grain boundary and interface energies ---$\gamma\GB$, $\gamma_1$, $\gamma_2$, and $\gamma\SiGe$--- in the fashion of Eq.~(\ref{def-theta_0}). However if the interface energy is strongly anisotropic the angle cannot change continuously, only some particular angles can exist~\cite{Wulff-1901}.
As the source of the faceting is not known, we will consider these two possibilities: (i) the angles are set by ratios of interface energies and (ii) the angles are independent of these ratios.

\begin{figure}
\centering
    \includegraphics[width=8.5cm]{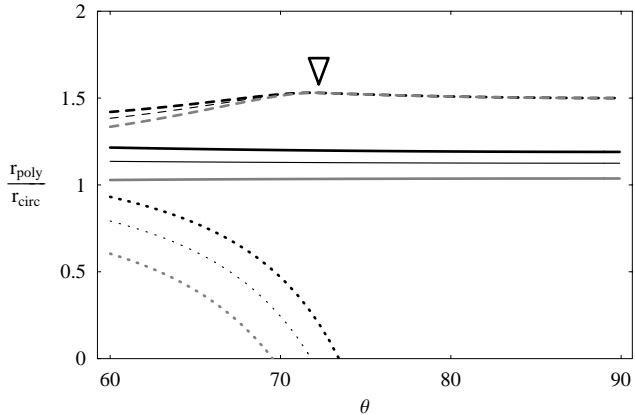}
\caption{\label{fig-ratio} The ratio of $r\facet$ to $r\iso$ as a function of the dihedral angle $\theta$. 
Dotted lines: $\gamma_1=0.95\gamma_2$; solid lines: $\gamma_1=\gamma_2$; dashed lines: $\gamma_1=1.05\gamma_2$. Thin black lines: $\theta_0=\theta$, thick black lines: $\cos \theta_0 = 1.1 \cos \theta$, thick gray lines: $\cos \theta_0 = 0.9 \cos \theta$. The arrow shows where $2\ell=w$.}
\end{figure}

If the angle $\theta$ for faceted films is set by the equilibrium between grain boundary and interface energies then $\theta=\theta_0$ (for rounded grains $\cos\theta_0 = {\gamma\GB}/{(2\gamma\I)}$ and $\cos\theta_0 = {\gamma\GB}/{(2\gamma_2)}$ for polygonal grains).
The thin black lines in Fig.~\ref{fig-ratio} show the value of $r\facet/r\iso$ as a function of the angle $\theta$ in this case.
A polygonal film is \emph{intrinsically} less stable ---it needs to be thicker than a rounded film to avoid agglomeration--- for purely geometric reasons (i.e.\ in the absence of interface energy anisotropy, solid line). If $\gamma_1>\gamma_2$, polygonal grains agglomerate even faster: a polygonal film needs to be about 50\% thicker than a rounded film to avoid agglomeration if $\gamma_1$ is larger than $\gamma_2$ by as little as 5\% (dashed line). If $\gamma_1<\gamma_2$ on the other hand, polygonal grains are more stable than rounded grains (dotted line). Even in the absence of strong anisotropy $\gamma_1$ may be different from $\gamma_2$ as they correspond to interfaces between different materials, MSi$_{1-u}$Ge$_u$--Si$_{1-x}$Ge$_x$ and MSi$_{1-u}$Ge$_u$--Si$_{1-z}$Ge$_z$ respectively.

If the angle $\theta$ is set by the anisotropy of the interface energy then $\theta$ needs not be equal to $\theta_0$. The thick lines in Fig.~\ref{fig-ratio} show $r\facet/r\iso$ as a function of the angle $\theta$ for different values of $\theta_0$.
The results are similar to the $\theta = \theta_0$ case: a polygonal film is less stable unless $\gamma_1<\gamma_2$ (dotted lines). The difference between $theta$ and $\theta_0$ is much less important than that between $\gamma_1$ and $\gamma_2$.

If minimizing the free energy leads to $2\ell > w$ one uses $2\ell = w$ instead to calculate $r\facet$. This is the case on the left of the arrow in Fig.~\ref{fig-ratio}. Grains are then ``pointy'' as can be seen for instance in Ref.~\onlinecite{Jarmar-JAP-05}.
On the other hand, if minimization of the energy leads to $\ell <0$,
$\ell$ is equal to zero and $r\facet = 0$. This is the case for the dotted lines in Fig.~\ref{fig-ratio} for large $\theta$. In such a case the film would not groove and {\itshape a fortiori} not agglomerate. Erbium disilicide films for instance do not exhibit any grooving. ErSi$_2$, which is hexagonal, grows epitaxially on Si(001) with its c-axis parallel to the substrate \cite{tan-06} but there probably is no epitaxial relationships for other directions due to the difference of crystal structure between film and substrate. Thus $\gamma_1$ is smaller than $\gamma_2$, which suppresses grooving.

We showed that strained substrates favor agglomeration and  inhomogeneities of the germanium composition in germanosilicides.
Grain shape too can affect agglomeration: polygonal films generally need to be thicker than rounded ones to avoid agglomerating. This result is independent of the cause of the faceting of the grains.

\vspace{\parsep}
I wish to thank Dongzhi Chi and his group for showing me and  discussing their experimental results.

\bibliography{../Pt-results/silicides}
\bibliographystyle{apsrev}

\end{document}